# Introducing Novel Droplet Generators: Enhanced Mixing Efficiency and Reduced Droplet Size

Ali Kheirkhah Barzoki[1,*]

[1] Department of Mechanical Engineering, Sharif University of Technology, Tehran, Iran

[*] Correspondence: ali_kheirkhahbarzoki@mech.sharif.edu

## Abstract

Nowadays, droplet microfluidics has become widely utilized for high-throughput assays. Efficient mixing is crucial for initiating biochemical reactions in many applications. Rapid mixing during droplet formation eliminates the need for incorporating micromixers, which can complicate the chip design. Furthermore, immediate mixing of substances upon contact can significantly improve the consistency of chemical reactions and resulting products. This study introduces three innovative designs for droplet generators that achieve efficient mixing and produce small droplets. The T-cross and cross-T geometries combine cross and T junction mixing mechanisms, resulting in improved mixing efficiency. Numerical simulations were conducted to compare these novel geometries with traditional T and cross junctions in terms of mixing index, droplet diameter, and eccentricity. The cross-T geometry exhibited the highest mixing index and produced the smallest droplets, increasing the mixing index by 10% compared to the T junction. While the T junction has the best mixing efficiency among traditional droplet generators, it produces larger droplets, which can increase the risk of contamination due to contact with the microchannel walls. Therefore, the cross-T geometry is highly desirable in most applications due to its production of considerably smaller droplets. Other new geometries also demonstrated comparable mixing efficiency to the T junction. The cross junction exhibited the lowest mixing efficiency and produced larger droplets compared to the cross-T geometry. Thus, the novel geometries, particularly the cross-T geometry, are a favorable choice for applications where both high mixing efficiency and small droplet sizes are important.

## Keywords

## 1. Introduction

The growing interest in chemistry and biology has amplified the demand for Lab-on-a-chip (LOC) design and processing technology. Microfluidics technology, a crucial aspect of biomedicine and biotechnology, has been widely utilized in various fields such as chemical synthesis, drug delivery, cell separation, and diagnostics (Farahinia et al., 2021; Kheirkhah Barzoki et al., 2023; Ma et al., 2022; Mao et al., 2020; Niculescu et al., 2021). This is due to its remarkable benefits in improving mass transfer, reducing the need for large sample and reagent quantities, and minimizing time and energy consumption (Battat et al., 2022).

Droplet-based microfluidic systems have gained significant attention within the field of microfluidics due to their ability to enable a wide range of functions. These systems have proven to be highly versatile, allowing for tasks such as biochemistry, single cell analysis, chemical synthesis, polymerase chain reaction (PCR), loop-mediated isothermal amplification (LAMP), and the production of specialized microparticles and nanoparticles (Amirifar et al., 2022; A. Jalili et al., 2021; Jiang et al., 2023; Oliveira et al., 2020; Sánchez Barea et al., 2019; Wang et al., 2020). The main benefit of droplet-based microfluidics is its ability to accurately encapsulate extremely small amounts of reaction components, ranging from femto- to nanoliters. This allows for quick mixing and minimal heat transfer within each droplet, leading to precise control over reaction conditions. Additionally, using droplets as reaction chambers with minimal variation in volume ensures a uniform environment for the reaction (Payne et al., 2020). In many applications involving microfluidic systems, efficient mixing is crucial to their performance. However, due to their small size, the flow in these systems is typically laminar, which limits mixing to diffusion. As a result, achieving adequate mixing can be challenging (Bahrami & Bayareh, 2022; X. Chen et al., 2021).

One method suggested to improve mixing in microfluidic devices is the use of micromixers, which can be either passive or active. Passive micromixers employ specific geometries or microstructures within the microfluidic channels to enhance the interaction between fluids and promote more efficient mixing (Rampalli et al., 2020; Raza et al., 2020; Shi, Huang, et al., 2021; Shi, Wang, et al., 2021; Tokas et al., 2021). In contrast, active micromixers utilize control elements that necessitate extra energy, such as electroosmotic, magnetic, and acoustofluidic mixing (Bahrami et al., 2022; Buglie & Tamrin, 2022; Z. Chen et al., 2022; H. Jalili et al., 2020; Mondal et al., 2021). While passive mixers face difficulties in achieving high mixing efficiency, active mixers offer the advantage of more efficient mixing of reaction components. However, this comes at the expense of increased energy consumption and a more intricate device design. Numerous studies have focused on enhancing the mixing index in droplet-based microfluidic systems by incorporating micromixers directly after the droplet generator (Ghazimirsaeed et al., 2021; Harshe et al., 2016; Madadelahi & Shamloo, 2017; Yu et al., 2022; Yu & Chen, 2022). Implementing micromixers induce advection in the droplets, thereby improving mixing. However, it is important to consider that incorporating micromixers adds complexity to the chip, increases its size, and restricts the frequency of droplet generation due to increased hydrodynamic resistance (Belousov et al., 2021). As a result, there is a growing need for alternative mixing methods that do not rely on micromixers.

An alternative approach in mixing in the microfluidic chips involves using droplets to mix reagents during the droplet formation process. When two immiscible fluids come together at a junction, different forces like viscosity, surface tension, and pressure gradient interact. As a result, the droplet separates from the dispersed phase and continues to flow downstream. By maintaining a controlled droplet size, this technique prevents the axial dispersion of mixing components, enabling rapid mixing through the internal circulation of the droplet (Burns & Ramshaw, 2001; Ward & Fan, 2015). Belousov et al. developed an asymmetric flow focusing droplet generator that enhances mixing during the droplet formation stage and demonstrated a six-fold increase in mixing speed compared to symmetric design. It's important to note that flow focusing has its own advantages. It operates by compressing the dispersed phase by the continuous phase from two side channels leading to the formation of droplets. This design ensures that droplets do not come into contact with the channel walls, minimizing any potential contamination in the dispersed phase components. However, the T junction offers the advantage of more efficient mixing although the droplet will be larger and more in contact with the walls of the channel. Therefore, it is reasonable to consider modifying the flow focusing design to further improve mixing capabilities.

In this research, I propose three novel droplet generators based on the conventional flow-focusing and T junction geometries. In comparison to the cross junction and T junction, the modifications in these geometries provide higher mixing efficiencies while keeping the droplet size relatively small. A comprehensive study is performed to compare the mixing efficiency, droplet diameter, and eccentricity of the droplets in these geometries. The proposed novel geometries entail asymmetric cross junction, T-cross junction, and cross-T junction (see **Fig. 1**). The simulations demonstrated that these novel designs, by combining the mixing mechanisms of the T junction and cross junction, provide improved mixing efficiency with droplet diameter smaller than that in the T junction geometry. This study offers valuable insights into the selection of optimal designs to improve mixing efficiency in droplet-based systems.

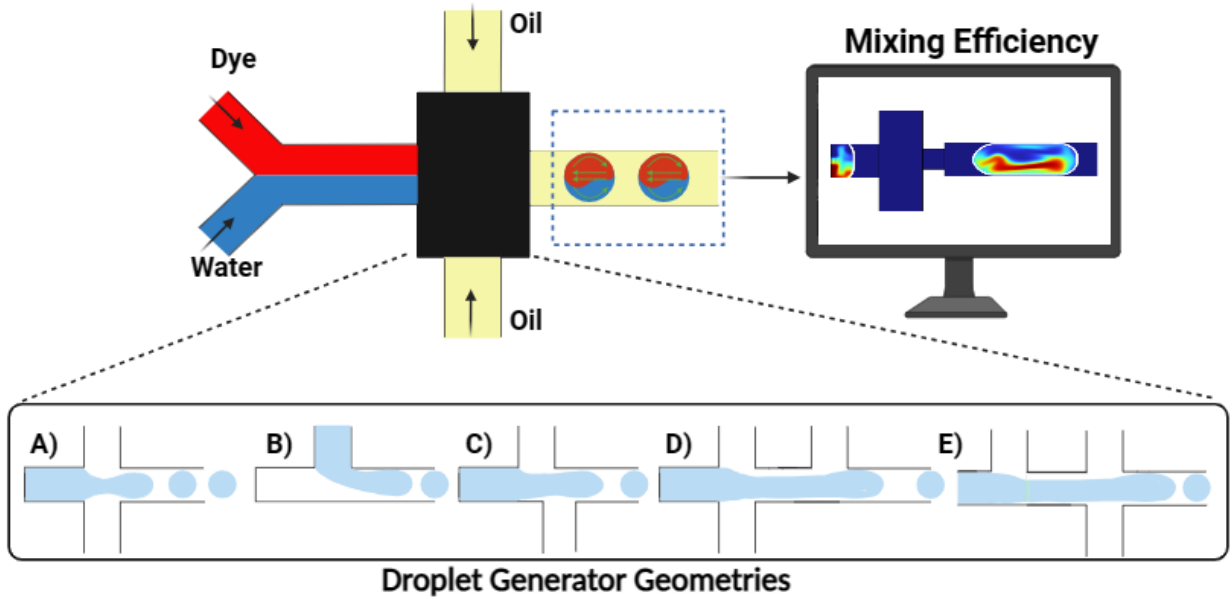

**Figure 1. The schematic illustration of the study.** There are five different geometries investigated: **A)** cross junction, **B)** T junction, **C)** asymmetric cross junction, **D)** cross-T junction, and **E)** T-cross junction.

## 2. Methods

### 2.1. Computational Method

Numerical simulations were conducted using the finite element method (FEM) to evaluate the mixing efficiency within droplets. Three sets of equations needed to be simultaneously solved in order to determine the mixing efficiency. Parallel Direct Sparse Solver (PARDISO) with a residual tolerance of 1E-3 was implemented to solve the fluid flow and phase-field equations. The mixing equation was solved using the Multifrontal Massively Parallel Sparse Direct Solver (MUMPS) with left preconditioning and a residual tolerance set at 1E-3. The discretization of pressure and velocity involved the use of first-order and second-order elements, respectively. Linear discretization was applied to both the phase-field and mixing solutions. At each time step, the Newton method was utilized to linearize the set of non-linear equations. Triangular (tri) elements were used to discretize the entire domain, including the boundary layers. For the simulations in this study, 2D geometries were employed. While 3D geometries may provide the most precise outcomes, 2D geometries have proven to be useful in providing valuable information and showing acceptable consistency with experimental data, all without the need for extensive computational power (Belousov et al., 2021).

### 2.2. Governing Equations

Navier-Stokes equations were used to model the liquid mediums. Considering incompressible fluid in the laminar flow regime, these governing equations were then simplified as in **Eqs. (1) and (2)**, as the maximum fluid flow rate was kept at 10 µL/min ($Re = \frac{\rho u w}{\mu} \approx 1$).

$$\nabla \cdot \vec{u} = 0 \tag{1}$$

$$\rho \frac{\partial \vec{u}}{\partial t} + \rho (\vec{u} \cdot \nabla) \vec{u} = -\nabla P + \mu \nabla^2 \vec{u} + \vec{F_s} \tag{2}$$

where $\vec{u}$ represents the velocity vector ($m/s$), P is the pressure ($Pa$), $\mu$ shows the dynamic viscosity ($Pa \cdot s$), $\rho$ denotes the fluid density ($kg/m^3$), and $F_s$ is the surface tension force (N).

In order to track the fluid-fluid interface in our two-phase system, the Cahn–Hilliard equation was utilized (Jacqmin, 1999):

$$\frac{\partial \phi}{\partial t} + \vec{u} \nabla \phi = \nabla \cdot \left( \frac{\gamma \lambda}{\varepsilon^2} \nabla \Psi \right) \tag{3}$$

$$\Psi = -\nabla \cdot (\varepsilon^2 \nabla \phi) + \phi(\phi^2 - 1) \tag{4}$$

In **Eqs. (3) and (4)**, $\phi$ represents the phase variable ensuring a smooth transition across the phase interface; $\varepsilon$ shows the interfacial thickness, which was set to half of the maximum mesh size; Ψ is an auxiliary variable in order to reduce the differential equation order. In **Eq (3)**, $\gamma$ denotes the mobility parameter, which is proportional to the square of interface thickness $\gamma = \chi\, \varepsilon^2$, where $\chi$ is the mobility tuning parameter, which was set to 1 in this study. $\lambda$ is the mixing energy density defined as $\lambda = {3\sigma}/{\sqrt{8}\varepsilon}$, where $\sigma$ is the surface tension coefficient. In **Eq. (2)**, by multiplying the chemical potential (G) by the gradient of the phase field ($F_s = G\nabla\phi$) the surface tension force could be determined. The definition of the chemical potential is as follows (Kim, 2012):

$$G = \lambda\left(-\nabla_t^2 \phi + \frac{\phi(\phi^2-1)}{\varepsilon^2}\right) \tag{5}$$

The main advantage of using the phase-field method to model two-phase flows is its capability to accurately calculate the displacement of the contact line while maintaining a no-slip boundary condition for fluid velocity. Moreover, it effectively eliminates pressure discontinuities at corners and prevents the formation of artificial vortices in intersecting channel regions. Additionally, as an interface capturing technique, it allows for precise resolution of droplet breakup.

In order to simulate the mixing process, the convective-diffusive mass transport equation was utilized:

$$\vec{u} \cdot \nabla c = D\nabla^2 c \tag{6}$$

where D and c are the diffusion coefficient ($m^2/s$) and the dye concentration ($mol/m^3$), respectively. To calculate the mixing index (MI), the dye concentration in each mesh cell was considered using the following formula (X. Chen et al., 2017):

$$MI\ (\%) = \left(1 - \sqrt{\frac{\iint (c-\bar{c})^2 dA}{A \cdot \bar{c}^2}}\right) \times 100 \tag{7}$$

where c, $\bar{c}$, and $c_{max}$ are the dye concentration, average and maximum dye concentration ($mol/m^3$) in the droplet. A denotes the droplet area ($m^2$).

It is worth mentioning that during this investigation, the characteristics of the dye solution were considered to be the same as those of DI water, which has a density of 1000 $kg/m^3$ and a dynamic viscosity of 1 $mPa.s$. Furthermore, the simulations used properties for olive oil, with a density of 917 $kg/m^3$ and a dynamic viscosity of 84 $mPa.s$. The surface tension was set at 74 $mN/m^2$. The boundary conditions used in the simulations can be found in Table 1, which includes $u_n, u_t, \mu$, V$_{interface}$, and Q representing the normal and tangential velocities, dynamic viscosity, velocity of the interface, and flow rate, respectively.

**Table 1:** Boundary conditions.

|  | *Boundary Condition* |
|---|---|
| *Liquid-Liquid Interface* | $u_n\|_{in} = u_n\|_{out} = V_{interface}$<br>$u_t\|_{in} = u_t\|_{out}$ |
| Liquid-Solid Interface | $\mu_{in}\frac{\partial u_t}{\partial r}\|_{in} = \mu_{out}\frac{\partial u_t}{\partial r}\|_{out}$ |
| | $u_{interface} = V_{interface}$ (no slip) |
| inlets | $Q = cte$ |
| Outlet | $P=0$ |

To have a meaningful comparison between different geometries, the total flow rate of the continuous phase was considered 10 $\mu l/min$. As a result, in geometries with two continuous phase inlets, like cross junction and asymmetric cross junction, each inlet has a flow rate of 5 $\mu l/min$, while in geometries with three continuous phase inlets, like T-cross and cross-T geometries, each inlet has a flow rate of 3.33 $\mu l/min$. In the T junction geometry, the continuous phase flow rate was considered 10 $\mu l/min$. The geometries and dimensions are illustrated in **Fig. 2**.

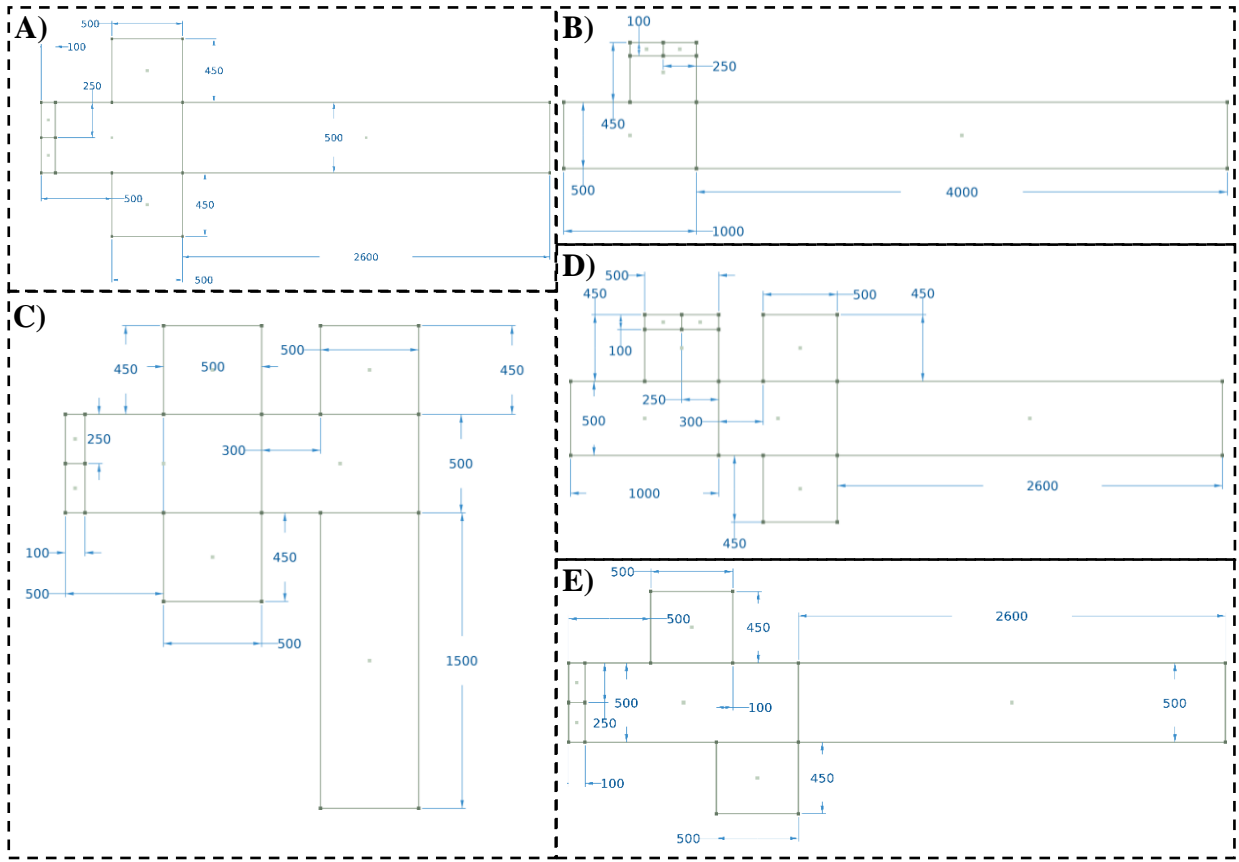

**Figure 2. Droplet generator geometries.** The dimensions of the **A)** cross junction, **B)** T junction, **C)** cross-T, **D)** T-cross, and **E)** asymmetric cross junction.

## 2.3. Model Validation

To verify the accuracy of the simulations, the model was validated by the results of the research by Belousov et al. (Belousov et al., 2021). The mixing index in different flow rate ratios of the dispersed phase to the continuous phase ($Q_d/Q_c$) were compared. The result is illustrated in **Fig. 3**. The asymmetric geometry with a depth of 40 $\mu m$ was considered for the validation. The simulations show a proper agreement with the results by Belousov et al. One of the simulations of the validation is depicted in the **Fig. 3b**.

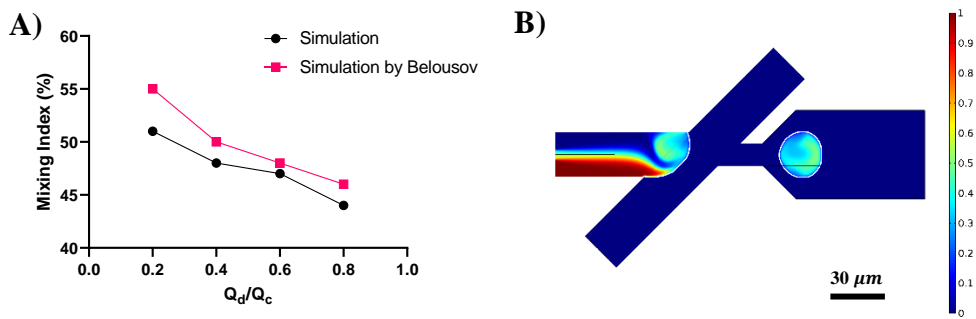

**Figure 3. Validation of the simulations in different flow rate ratios. A)** Comparative analysis of the mixing index and **B)** Simulation of the droplet formation and mixing for the flow rate of 0.2. The continuous phase flow rate ($Q_c$) is 1 $\mu l/min$.

## 3. Results and Discussion
### 3.1. Fluid Dynamics

**Fig. 2** demonstrates the geometries of droplet generators investigated in this research. In addition to the two well-known geometries, T junction and cross junction, three novel geometries are introduced in this study. The main idea behind these novel designs is combining the mixing mechanism and the recirculation vortex types of the cross junction and the T junction. **Fig. 4** illustrates that during the filling stage, the velocity of the interface is much lower than that of the continuous phase, resulting in the creation of one or two fluid recirculation vortices in the dispersed phase. These vortices are induced by the flow of the continuous phase due to the boundary conditions at the liquid-liquid interface. In the T junction, there is one large recirculation vortex during the filling stage, which mixes the species throughout the droplet, resulting in higher mixing efficiency compared to the cross junction. Conversely, in the cross junction, two symmetrical vortices exist, but they do not significantly contribute to the mixing process as they only mix the species in the two upper and lower halves of the droplet separately.

The T-cross (**Fig. 4C**) and cross-T (**Fig. 4D**) geometries take advantage of the mixing mechanisms of both the T junction and the cross junction. It is important to note that in the T junction, the vortex effectively mixes all the components inside the droplet (**Fig. 4B**). However, the velocity near the center of the vortex approaches zero, resulting in poor mixing efficiency in this area. On the other hand, in the cross junction, the two vortices in the droplet can mix the reagents in the center of the droplet and its sides by circulating the species from center to sides and vice versa (**Fig. 4A**). As a result, the two types of vortices in the T junction and the cross junction can complement each other's mixing performance, leading to higher mixing efficiency in the T-cross and cross-T geometries compared to the typical T and cross junctions.

In the asymmetric cross junction geometry (**Fig. 4E**), the offset between the two inlets of the continuous phase causes a delay in the formation of the two counter-circulating vortices in the cross junction. As the dispersed phase approaches the first inlet of the continuous phase, a single vortex forms inside the dispersed phase, similar to the T junction geometry. However, upon reaching the second continuous phase inlet, another vortex emerges within the dispersed phase with an opposite direction. At this point, the mixing mechanism becomes similar to that of the cross junction, as there are now two recirculation vortices. Therefore, it can be said that the asymmetric cross junction behaves similarly to the T-cross geometry.

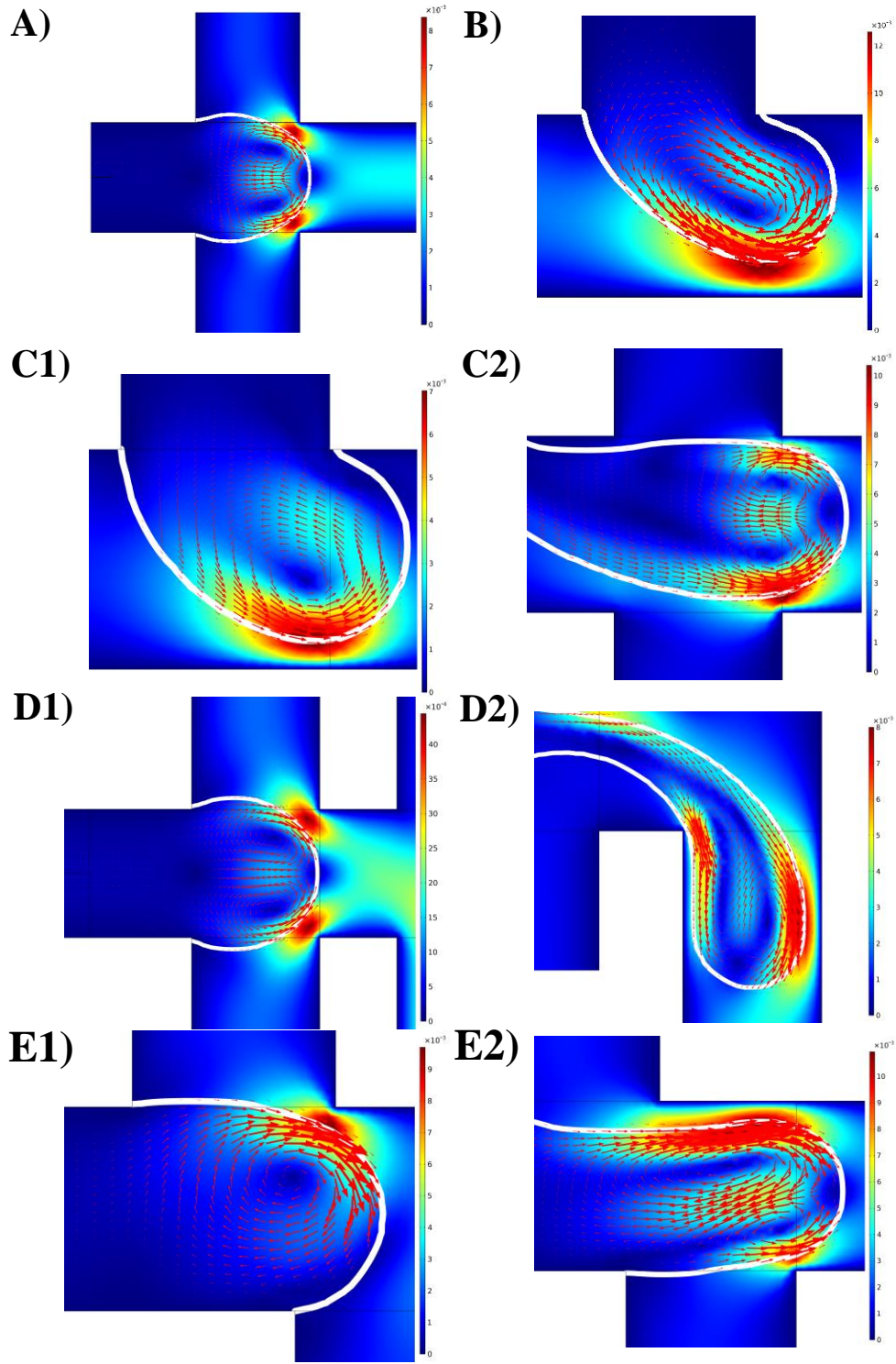

**Figure 4. 2D simulation of the droplet formation.** Velocity distribution during the filling stage in **A)** cross junction, **B)** T junction, **C1)** T junction of the T-cross, **C2)** cross junction of the T-cross, **D1)** cross junction of the cross-T, **D2)** T junction of the cross-T, **E1)** first junction of the asymmetric cross junction, and **E2)** second junction of the asymmetric cross junction. The color bars show the velocity magnitude in m/s.

## 3.2. Droplet Diameter and Eccentricity

In many applications of droplet-based microfluidics, it is crucial to keep the created droplets away from the walls of the channels to avoid contamination of the droplets and their contents. Therefore, there is a high demand for producing small droplets. In addition to efficient mixing inside the droplets, the size and shape of the droplets are important factors to consider when comparing different droplet-generator designs.

**Fig. 5** illustrates the variations of droplet diameter and eccentricity as a function of the flow rate ratio of the dispersed phase to the continuous phase in different geometries. In this study, by considering droplets as ellipses, eccentricity was defined as $e = \sqrt{1 - b^2/a^2}$, where a and b represent the longest and the shortest diameters of the ellipse, respectively. The eccentricity value is used to gauge how much the droplet's shape differs from a perfect circle. Almost in all geometries, there is a consistent rise in the droplets' eccentricity as the flow rate of the dispersed phase increases. This happens because the larger droplet size exceeds the channel's width capacity, causing it to deform from its natural spherical shape. To be noted that the droplet diameter was calculated by considering the droplet as a circle.

When comparing these geometries, it becomes evident that the T-cross junction geometry exhibits the highest values for both droplet size and eccentricity, while the smallest values are observed in the cross junction and cross-T junction geometries. The novel geometries, asymmetric cross junction and cross-T junction, offer relatively small diameters in comparison to the T junction and comparable to the cross junction, which can be beneficial in many of applications. The T-cross junction produce the largest droplets. This is because the first junction in this geometry, T junction, is mainly responsible for the size of the droplet and the main shear force from the continuous phase is exerted on the dispersed phase in this junction. The second junction in this geometry, cross junction, has a more influential effect on the mixing efficiency since it determines the last recirculation vortex type inside the droplet.

In the contrary, in the cross-T junction, the first junction is the cross junction. As we expected, the produced droplets in this geometry are smaller than the T-cross junction. The second junction in this geometry, the T junction, makes a recirculation vortex as in the conventional T junction and increases the mixing efficiency.

In the asymmetric cross geometry, the pinch-off process and the droplet formation is similar to that in the conventional cross junction. Therefore, we expect the droplet size in this geometry to be around that in the cross junction. However, the offset between the two inlets of the continuous phase, increases the mixing efficiency inside the droplet due to the delay in the formation of the vortices inside droplets. Considering the droplet size and eccentricity in all geometries, the two novel geometries, cross-T junction and asymmetric cross junction, produce droplets with small sizes comparable to the conventional cross junction.

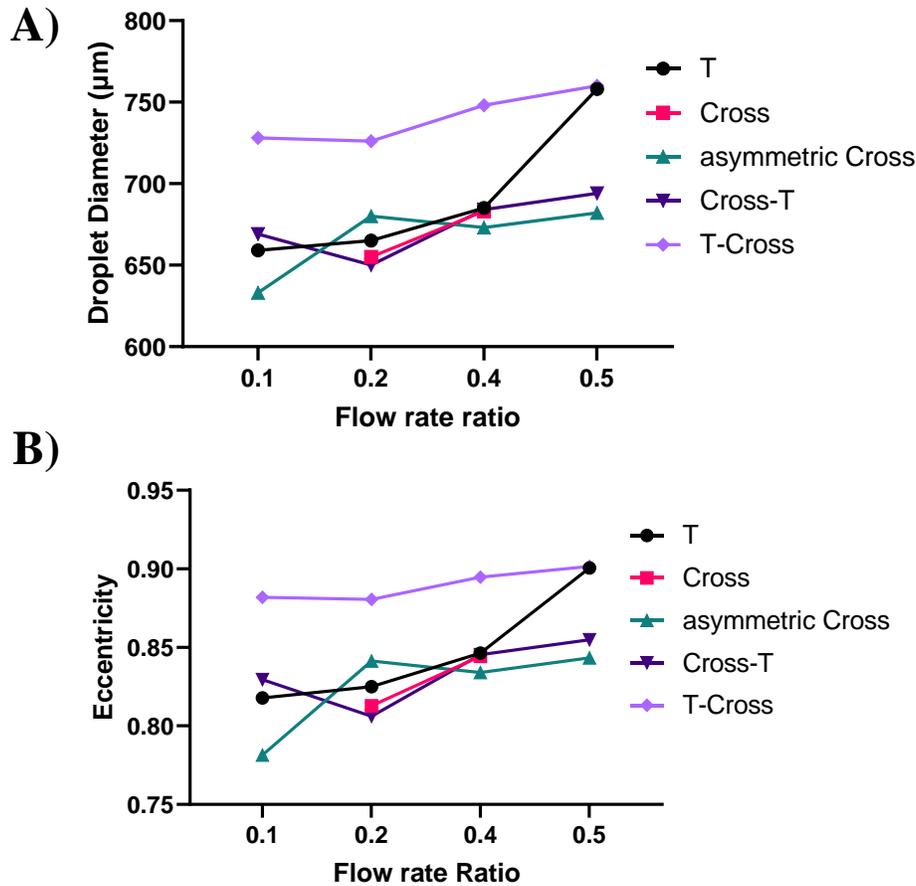

**Figure 5. Diameter and eccentricity variations of droplets with flow rate ratio.** Comparative analysis of **A)** droplet diameter and **B)** eccentricity in various geometries. The total flow rate of the continuous phase is 10 $\mu l/min$.

### 3.3. Mixing Efficiency

The mixing efficiency is important in many applications, and micromixers are commonly used in microfluidic systems to ensure proper mixing of substances within droplets. However, incorporating micromixers may require extra space. Additionally, instant mixing of substances upon contact can greatly enhance the consistency of chemical reactions and their resulting products. Given these factors, using a droplet generator that can mix substances within the droplet during the filling stage seems like a promising alternative to using micromixers.

**Fig. 6** presents an evaluation of the efficiency of mixing within droplets immediately after their formation in various droplet generators. It is important to note that the process of droplet formation and the characteristics of recirculation vortices inside the droplets significantly influence the mixing efficiency. Almost in all geometries, a consistent trend is observed: as the flow rate of the dispersed phase increases, the mixing efficiency decreases. This phenomenon can be explained by the reduction in filling stage duration that occurs when the dispersed phase flow rate is increased. This implies that the substances inside the droplets do not have enough time for thorough mixing due to the presence of formed vortices. However, in asymmetric cross geometry, the trend is reverse.

The cross-T junction geometry exhibits the highest mixing efficiency due to the existence of both types of vortices: two counter-circulating in the cross junction and one vortex in the T junction of this geometry. These two types of vortices complement each other, leading to a high mixing efficiency. This innovative geometry increases the mixing efficiency by approximately 10% compared to the conventional T junction, which offers the best mixing efficiency among conventional droplet generators. It is worth noting that there is an increase in the mixing efficiency while the size of droplet in the cross-T is smaller than in the T junction. Other innovative geometries also demonstrate comparable mixing efficiency compared to the T junction and cross junction. The cross junction exhibits the lowest mixing efficiency due to the symmetry in this geometry, where the two flows of the continuous phase are identical. As a result, two identical vortices but in opposite directions are formed inside the droplet, which only mix the substances in the upper and lower halves of the droplet separately. Because there is no single vortex enveloping the entire droplet, this unique configuration leads to the lowest mixing efficiency values among all the geometries.

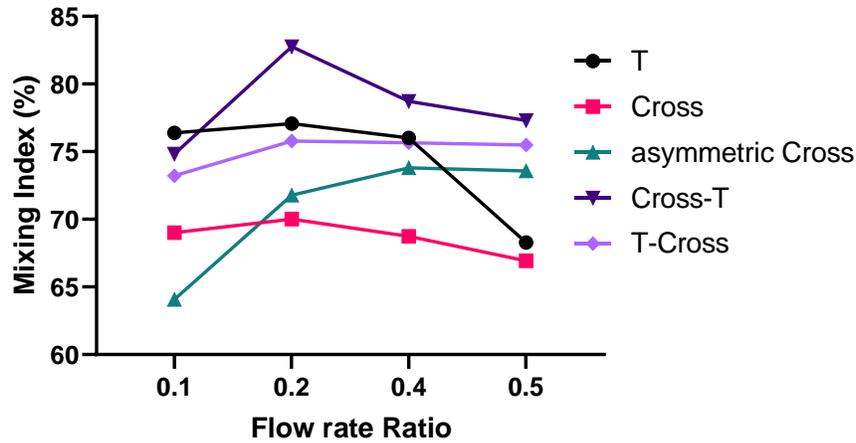

**Figure 6. Mixing efficiency variations within droplets with flow rate ratio in different geometries**. Comparative analysis of mixing in various Geometries. The overall flow rate of the continuous phase is $10 \, \mu l/min$.

## 4. Conclusions

In this research, three novel droplet generator designs that offer efficient mixing and small droplet sizes are introduced: T-cross, cross-T, and asymmetric cross geometries. The T-cross and cross-T geometries combine cross and T junction mixing mechanisms, resulting in increased mixing efficiency. Numerical simulations were conducted and these novel geometries were compared with traditional T and cross junctions in terms of mixing index, droplet diameter, and eccentricity. In all cases, the overall flow rate of the continuous phase was considered the same ($10 \, \mu l/min$) to have identical conditions for all geometries. The cross-T geometry had the highest mixing index and produced the smallest droplets, increasing the mixing index by 10% compared to the T junction. While the T junction has the best mixing efficiency among traditional droplet generators, it produces larger droplets, which can increase the risk of contamination due to the contact of the droplet with walls of the microchannel. The cross-T geometry is therefore highly desirable in most applications since produces droplets with considerably small droplets. Other new geometries also showed comparable mixing efficiency to the T junction, with droplet sizes between those of the cross and T junctions. The cross junction had the lowest mixing efficiency and droplet diameter larger than the cross-T geometry. Therefore, the novel geometries, particularly the cross-T geometry, are a favorable choice for applications where both high mixing efficiency and smaller droplet sizes are important.

In the future, I plan to investigate the impact of offset size on the mixing index in the asymmetric cross junction. Additionally, studying the distance between the T and cross junctions in the T-cross and cross-T geometries would be beneficial.

17